\documentclass[prd,aps,floats,twocolumn]{revtex4}
\usepackage{amssymb}
\usepackage[dvips]{graphicx}
\begin{document}
\newcommand{\beq}{\begin{equation}}
\newcommand{\eeq}{\end{equation}}
\newcommand{\ie}{{\sl i.e\/}}
\newcommand{\half}{\frac 1 2}
\newcommand{\lag}{\cal L}
\newcommand{\ove}{\overline}
\newcommand{\et}{{\em et al}}
\newcommand{\Prd}{Phys. Rev D}
\newcommand{\Prl}{Phys. Rev. Lett.}
\newcommand{\Plb}{Phys. Lett. B}
\newcommand{\Cqg}{Class. Quantum Grav.}
\newcommand{\Grg}{Grav....}
\newcommand{\Np}{Nuc. Phys.}
\newcommand{\Fp}{Found. Phys.}
\renewcommand{\baselinestretch}{1.2}

\title{Cyclic Magnetic Universe }

\author{M. Novello, Aline N. Araujo and J. M. Salim}
\affiliation{Institute of Cosmology, Relativity and Astrophysics (ICRA/CBPF) \\
 Rua Dr. Xavier Sigaud, 150, CEP 22290-180, Rio de Janeiro, Brazil}

\date{\today}

\begin{abstract}

Recent works have shown the important role Nonlinear Electrodynamics
(NLED) can have in two crucial questions of Cosmology, concerning
particular moments of its evolution for very large and for
low-curvature regimes, that is for very condensed phase and at the
present period of acceleration. We present here a toy model of a
complete cosmological scenario in which the main factor responsible
for the geometry is a nonlinear magnetic field which produces a FRW
homogeneous and isotropic geometry. In this scenario we distinguish
four distinct phases: a bouncing period, a radiation era, an
acceleration era and a re-bouncing. It has
 already been shown that in NLED a strong magnetic field can
overcome the inevitability of a singular region typical of linear
Maxwell theory; on the other extreme situation, that is for very
weak magnetic field it can accelerate the expansion. The present
model goes one step further: after the acceleration phase the
universe re-bounces and enter in a collapse era. This behavior is
a manifestation of the invariance under the dual map of the scale
factor $ a(t) \rightarrow 1/ a(t),$ a consequence of the
corresponding inverse symmetry of the electromagnetic field ($ F
\rightarrow 1/ F,$ where $F \equiv F^{\mu\nu}F_{\mu\nu}$) of the
NLED theory presented here. Such sequence
collapse-bouncing-expansion-acceleration-re-bouncing-collapse
constitutes a basic unitary element for the structure of the
universe that can be repeated indefinitely yielding what we call a
Cyclic Magnetic Universe.

\end{abstract}

\vskip2pc
 \maketitle

\section{Introduction}

 In the last years there has been increasing of interest on
 the cosmological effects induced by Nonlinear Electrodynamics
 (NLED) \cite{novello1}.  The main reason for this is related to the drastic
 modification NLED provokes in the behavior of the cosmological
 geometry in respect to two of the most important questions of
 standard cosmology, that is, the initial singularity and the
 acceleration of the scale factor. Indeed, NLED provides
 worthwhile alternatives to solve these two problems in a
 unified way, that is without invoking different mechanisms for
 each one of them separately. Such economy of hypotheses is certainly
 welcome.  The partial analysis of each one of these problems was initiated in
  \cite{novello1,novello2}. Here we will present a new cosmological model, that unifies both descriptions.

The general form for the dynamics of the electromagnetic
field, compatible with covariance and gauge conservation
principles reduces to  $L = L(F),$ where $F \equiv F^{\mu\nu}
F_{\mu\nu}.$ We do not consider here the other invariant $G \equiv
F^{\mu\nu} F^{*}_{\mu\nu}$ constructed with the dual, since its practical
importance disappears in cosmological framework once in our
scenario the average of the electric field vanishes in a magnetic
universe as we shall see in the next sections. Thus, the
Lagrangian appears as a regular function that can be developed as
positive or negative powers of the invariant $F.$ Positive powers
dominate the dynamics of the gravitational field in the
neighborhood of its moment of extremely high curvatures
\cite{novello1}. Negative powers control the other extreme, that
is, in the case of very weak electromagnetic fields
\cite{novello2}. In this case as it was pointed out previously it
modifies the evolution of the cosmic geometry for large values of
the scale factor, inducing the phenomenon of acceleration of the
universe. The arguments presented in \cite{lemoine}  make it worth
 considering that only the averaged magnetic field survives in a FRW
spatially homogeneous and isotropic geometry. Such configuration
of pure averaged magnetic field combined with the dynamic
equations of General Relativity received the generic name of
Magnetic Universe \cite{novello3}.

The most remarkable property of a Magnetic Universe configuration
is the fact that from the energy conservation law it follows that
the dependence on time of the magnetic field $H(t)$ is the same
irrespective of the specific form of the Lagrangian. This property
allows us to obtain the dependence of the magnetic field on the
scale factor $a(t),$ without knowing the particular form of the
Lagrangian $L(F).$ Indeed, as we will show later on, from the
energy-momentum conservation law it follows that $ H = H_{0} \,
a^{-2}.$ This dependence is responsible for the property which
states that strong magnetic fields dominates the geometry for
small values of the scale factor; on the other hand, weak fields
determines the evolution of the geometry for latter eras when the
radius is big enough to excite these terms.

In order to combine both effects, here we will analyze a toy
model. The symmetric behavior of the magnetic field in both
extremes -- that is for very strong and very weak regimes --
allows the appearance of a repetitive configuration of the kind
exhibited by an eternal cyclic universe.

Negative power of the field in the Lagrangian of the gravitational
field was used in \cite{carroll1} attempting to explain the
acceleration of the scale factor of the universe by modification of
the dynamics of the gravitational field by adding to the
Einstein-Hilbert action a term that depends on negative power of the
curvature, that is
$$
S =\frac{M_{\rm Pl}^2}{2} \int \sqrt{-g} \left( R -
\frac{\alpha ^4}{ R} \right) d^4x,
$$
This modification introduced an idea that is worth to be
generalized: the dynamics should be invariant with respect to the
inverse symmetry transformation. In other words,if $\mathbb{X}$
represents the invariant used to construct a Lagrangian for a given
field, the Action should be invariant under the map $\mathbb{X}
\rightarrow 1/\mathbb{X}.$ Since the Electrodynamics is the paradigm
of field theory, one should start the exam of such a principle into
the realm of this theory. In other words we will deal here with a
new symmetry between strong and weak electromagnetic field. In
\cite{novello2}, a model assuming this idea was presented and its
cosmological consequences analyzed. In this model, the action for
the electromagnetic field was modified by the addition of a new
term, namely \beq S = \int \sqrt{-g} \left( - \frac F 4+ \frac
\gamma F \right) d^4x. \label{action} \eeq This action yields an
accelerated expansion phase for the evolution of the universe, and
correctly describes the electric field of an isolated charge for a
sufficiently small value of parameter $\gamma.$ The acceleration
becomes a consequence of the properties of this dynamics for the
situation in which the field is weak.

In another cosmological context, in the strong regime, it has been
pointed out in the literature \cite{novello1} that NLED can produces
a bouncing, altering another important issue in Cosmology: the
singularity problem. In this article we would like to combine both
effects improving the action given in Eqn.(\ref{action}) to discuss
the consequences of NLED for both, weak and strong fields.

It is a well-known fact that under certain assumptions, the
standard cosmological model unavoidably leads to a singular
behavior of the curvature invariants in what has been termed the
Big Bang. This is a highly distressing state of affairs, because
in the presence of a singularity we are obliged to abandon the
rational description of Nature. It is possible that a complete
quantum cosmology could describe the state of affairs in a very
different and more complete way. For the time being, while such
complete quantum theory is not yet known, one should attempt to
explore alternatives that are allowed and that provide some sort
of phenomenological consequences of a more profound theory.

It is tempting then to investigate how NLED can give origin to an
unified scenario that not only accelerates the universe for weak
fields (latter cosmological era) but that is
also capable of avoiding an initial singularity as a consequence
of its properties in the strong regime.

Scenarios that avoid an initial singularity have been intensely
studied over the years. As an example of some latest realizations we
can mention the pre-big-bang universe \cite{pbb} and the ekpyrotic
universe \cite{ekpy}.  While these models are based on deep
modifications on conventional physics, that are extremely difficult
to be observed, the model we present here relies instead on the
electromagnetic field. The new ingredient that we introduce concerns
the dynamics that is rather different from that of Maxwell in
distinct regimes. Specifically, the Lagrangian we will work with is
given by \beq L_{T} = \alpha^{2} \, F^{2} -\frac{1}{4} \, F -
\frac{\mu^{2}}{F} + \frac{\beta^{2}}{F^{2}}. \label{lag} \eeq The
dimensional constants $\alpha, \beta$ and $\mu$ are to be determined
by observation. Thus the complete dynamics of electromagnetic and
gravitational fields are governed by Einstein equations plus $
L_{T}.$

We shall see that in Friedmann-Robertson-Walker (FRW) geometry we
can distinguish four typical eras which generate a basic unity of
the cosmos (BUC)
that repeat indefinitely.


The whole cosmological scenario is controlled by the energy density
$\rho$ and the pressure $p$ of the magnetic field. Each era of the
BUC is associated with a specific term of the Lagrangian. As we
shall see the conservation of the energy-momentum tensor implies
that the field dependence on the scale factor yields that the
invariant $ F $ is proportional to $ a^{- \, 4}.$ This dependence is
responsible by the different dominance of each term of the
Lagrangian in different phases. The first term $\alpha^{2} F^{2}$
dominates in very early epochs allowing a bouncing to avoid the
presence of a singularity. Let us call this the
\emph{\textbf{bouncing era}}. The second term is the Maxwell linear
action which dominates in the \emph{\textbf{radiation era}}. The
inverse term $\mu^{2} / F$ dominates in the
\emph{\textbf{acceleration era}}. Finally the last term $\beta^{2}
/F^{2}$ is responsible for a \emph{\textbf{re-bouncing}}. Thus each
 BUC can be described in the following way:

\begin{itemize}
\item{The bouncing era: There exists a collapsing phase that attains a minimum value
for the scale factor $a_{B}(t);$}
\item{The radiation era: after the bouncing,   $ \rho + 3 p$
changes the sign; the universe stops its acceleration and start
expanding with $\ddot{a} <0;$}
\item{The acceleration era: when the $1/F$ factor dominates the universe enters
an accelerated regime; }
\item{The re-bouncing era: when the term $1/F^{2}$ dominates, the acceleration
changes the sign and starts a phase in which $\ddot{a} <0$ once
more; the scale factor attains a maximum and re-bounces}
\end{itemize}
The universe starts a collapsing phase entering a new bouncing era.
This unity of four stages, the BUC, constitutes an eternal cyclic
configuration that repeats itself indefinitely.

The plan of the article is as follows. In section II we review the
Tolman process of average in order to conciliate the energy
distribution of the electromagnetic field with a spatially isotropic
geometry. Section III presents the notion of the Magnetic Universe
and its generic features concerning the dynamics of electromagnetic
field generated by a Lagrangian $L = L(F).$ Section IV presents the
conditions of bouncing and acceleration of a FRW universe in terms
of properties to be satisfied by $L.$ In section V we introduce the
notion of inverse symmetry of the electromagnetic field in a
cosmological context. This principle is used to complete the form of
the Lagrangian that guides the combined dynamics of the unique
 long-range fields yielding a spatially homogeneous and isotropic
nonsingular universe. In sections VI and VII we present a complete
scenario consisting of the four eras: a bouncing, an expansion
with negative acceleration, an accelerated phase and a
re-bouncing. We end with some comments on the form of the scale
factor and future developments. In appendix we present the
compatibility of our Lagrangian with standard Coulomb law and the
modifications induced on causal properties of nonlinear
electrodynamics.

\section{The average
procedure and the fluid representation}

The effects of a nonlinear electromagnetic theory in a
cosmological setting have been studied in several articles
\cite{portuga1}.

 Given a generic gauge-independent Lagrangian $L = L(F)$,
written in terms of the invariant $F \equiv F_{\mu\nu} F^{\mu\nu}$
it follows that the associated energy-momentum tensor, defined by
\begin{equation}
T_{\mu\nu} = \frac{2}{\sqrt {-\gamma}} \frac{\delta L
\sqrt{-\gamma}}{\delta \gamma^{\mu\nu}}, \label{n10}
\label{NL2}
\end{equation}
reduces to
\begin{equation}
T_{\mu\nu} = -4 \, L_{F} \, F_{\mu}{}^{\alpha} \, F_{\alpha\nu} -
L  \, g_{\mu\nu}. \label{NL3}\end{equation}
 In the standard cosmological scenario the metric
structure of space-time is provided by the FLRW geometry. For
compatibility with the cosmological framework, that is, in order
that an electromagnetic field can generates a homogeneous and
isotropic geometry an average procedure must be used. We define
the volumetric spatial average of a quantity $X$ at the time $t$
by \beq \overline X \equiv \lim_{V\rightarrow V_0} \frac 1 V \int
X \sqrt{-g}\;d^3x, \eeq where $V = \int \sqrt{-g}\;d^3x$ and $V_0$
is a sufficiently large time-dependent three-volume. In this
notation, for the electromagnetic field to act as a source for the
FLRW model we need to impose that \beq \overline E_i =0, \;\;\;
\overline H_i =0,\;\;\; \overline{E_i H_j}=0, \eeq \beq
\overline{E_iE_j}=-\frac 1 3 E^2 g_{ij}, \;\;\;\overline{H_iH_j} =
-\frac 1 3 H^2 g_{ij}. \eeq With these conditions, the
energy-momentum tensor of the EM field associated to  $L = L(F)$
can be written as that of a perfect fluid,
\begin{equation}
T_{\mu\nu} = (\rho + p) v_\mu v_\nu - p\; g_{\mu\nu}, \label{NL33}
\end{equation}
 where
\begin{eqnarray}
\rho &=& - L  - 4 L_{F} E^{2},  \nonumber \\
p &=& L  - \frac{4}{3} \, ( 2 H^{2} - E^{2}) \, L_{F}, \label{NL4}
\end{eqnarray}
and $L_F\equiv dL / dF.$

\section{Magnetic universe}

A particularly interesting case occurs when  only the average of
the magnetic part does not vanishes and  $E^{2} =0.$ Such
situation has been investigated in the cosmological framework
yielding what has been called \emph{magnetic universe.} This
should be a real possibility in the case of cosmology, since in
the early universe the electric field is screened by the charged
primordial plasma, while the magnetic field lines are frozen
\cite{lemoine}. In spite of this fact, in \cite{novello2}  some
attention was devoted to the mathematically interesting case in
which $E^2 = \sigma^{2} H^{2} \neq 0.$

An interesting feature of such magnetic universe comes from the
fact that it can be associated with a four-component
non-interacting perfect fluid. Let us give a brief proof of the
statement that in the cosmological context the energy-content that
follows from this theory can be described in terms of a perfect
fluid. We work with the standard form of the FLRW geometry in
Gaussian coordinates provided by (we limit the present analysis to
the Euclidean section)
\begin{equation}
ds^{2} = dt^{2} - a(t)^{2} \, \left( dr^{2} + r^{2} d\Omega^{2}
\right). \label{dez181}
\end{equation}

The expansion factor, $\theta$  defined as the divergence of the
fluid velocity reduces, in the present case, to the derivative of
logarithm of the scale factor
\begin{equation}
\theta \equiv v^{\mu}_{;\mu} = 3 \, \frac{\dot{a}}{a}
\label{dez182}
\end{equation}
The conservation of the energy-momentum tensor projected in the
direction of the co-moving velocity $v^{\mu} = \delta^{\mu}_{0}$
yields
\begin{equation}
\dot{\rho} + (\rho + p) \theta = 0 \label{M1}
\end{equation}
Using Lagrangian $L_{T}$ in the case of the magnetic universe
yields for the density of energy and pressure given in equations
(\ref{NL4}):
\begin{equation}
\rho = - \, \alpha^{2}\, F^{2} + \frac{1}{4} \, F +  \,
\frac{\mu^{2}}{F} -  \, \frac{\beta^{2}}{F^{2}} \label{19dez2}
\end{equation}
\begin{equation}
p = - \,\frac{5 \alpha^{2}}{3}\, F^{2} + \frac{1}{12} \, F - \frac{7
\mu^{2}}{3} \, \frac{1}{F} + \frac{11 \beta^{2}}{3} \,
\frac{1}{F^{2}} \label{19dez2}
\end{equation}
where
\begin{equation}
F=2H^2
\end{equation}
Substituting these values in the conservation law, it follows
\begin{equation}
L_{F} \left[(H^{2})\dot{} +  4 \, H^{2} \,
\frac{\dot{a}}{a}\right] = 0.
 \label{M2}
 \end{equation}
where $ L_{F}\equiv \partial L / \partial F.$

 The important result that follows from this equation is that
the dependence on the specific form of the Lagrangian appears as a
multiplicative factor. This property shows that any Lagrangian $
L(F)$ yields the same dependence of the field on the scale factor
irrespective of the particular form of the Lagrangian. Indeed,
equation (\ref{M2}) yields
\begin{equation}
 H = H_{0} \, a^{-2}.
\label{dez183}
\end{equation}
This property implies that for each power $F^{k}$ it is possible
to associate a specific fluid configuration with density of energy
$\rho_{k}$ and pressure $p_{k}$ in such a way that the
corresponding equation of state is given by \beq p_k = \left(
\frac{4k}{3} - 1 \right) \rho_k. \label{partialp} \eeq
 We restrict our analysis in
the present paper to the theory provided by a toy-model described
by the Lagrangian
\begin{eqnarray}
L_{T} &=& L_{1} + L_{2} + L_{3} + L_{4} \nonumber \\
&=&   \alpha^{2}\, F^{2} - \frac{1}{4}\, F  -  \,
\frac{\mu^{2}}{F} +  \, \frac{\beta^{2}}{F^{2}} \label{19dez1}
\end{eqnarray}
where $\alpha, \beta, \mu$ are parameters characterizing a concrete
specific model. For latter use we present the corresponding
many-fluid component associated to Lagrangian $L_{T}.$ We set for
the total density and pressure $ \rho_{T} = \sum \, \rho_{i}$ and $
p_{T} = \sum \, p_{i}$ where
\begin{eqnarray}
\rho_{1} &=& - \alpha^{2} \, F^{2} \,\, , \,\, p_{1} = \frac{5}{3} \rho_{1} \nonumber \\
\rho_{2} &=& \frac{1}{4} \, F  \,\,, \,\, p_{2} = \frac{1}{3}
\rho_{2} \nonumber \\
\rho_{3} &=&  \, \frac{\mu^{2}}{F} \,\, , \,\, p_{3} = - \, \frac{7}{3} \, \rho_{3} \nonumber \\
\rho_{4} &=& - \,  \frac{\beta^{2}}{F^{2}} \,\, , \,\,  p_{4} =
 - \, \frac{11}{3} \, \rho_{4}.
\end{eqnarray}
Or, using the dependence of the field on the scale factor equation
(\ref{dez183}),
\begin{eqnarray}
\rho_{1} &=& - \,  4 \alpha^{2}\, H_{0}^{4} \,   \frac{1}{a^{8}} \nonumber \\
\rho_{2} &=& \frac{H_{0}}{2} \,  \frac{1}{a^{4}} \nonumber \\
\rho_{3} &=&  \, \frac{\mu^{2}}{2 H_{0}^{2}} \, a^{4} \nonumber \\
\rho_{4} &=& - \,  \frac{\beta^{2}}{4 H_{0}^{4}} \, a^{8}.
\end{eqnarray}
Let us point out a remarkable property of the combined system of
this NLED generated by $L_{T}$ and Friedman equations of
cosmological evolution. A simple look into the above expressions
for the values of the density of energy exhibits what could be a
possible difficulty of this system in two extreme situations, that
is, when $F^{2}$ and $1/F^{2}$ terms dominate, since if the radius
of the universe can attain arbitrary small and/or arbitrary big
values, then one should face the question regarding the positivity
of its energy content. However, as we shall show in the next
sections, the combined system of equations of the cosmic metric
and the magnetic field described by General Relativity and NLED,
are such that a beautiful conspiracy occurs in such a way that the
negative contributions for the energy density that came from terms
$ L_{1}$ and $L_{4}$ never overcomes the positive terms that come
from $ L_{2}$ and $L_{3}.$ Before arriving at the undesirable
values where the density of energy could attain negative values,
the universe bounces ( for very large values of the field) and
re-bounces (in the other extreme, that is, for very small values)
to precisely avoid this difficulty. This occurs at the limit value
$\rho_{B} = \rho_{RB} = 0,$ as follows from equation
\begin{equation} \rho = \frac{\theta^{2}}{3}.
\label{29set11}
\end{equation}

We emphasize that this is not an extra condition imposed by hand
but a direct consequence of the dynamics described by $L_{T}.$
Indeed, at early stages of the expansion phase the dynamics is
controlled by the approximation Lagrangian $ L_{T} \approx L_{1,2}
= L_{1} + L_{2}$. Then
$$ \rho = \frac{F}{4} \, ( 1 - 4 \alpha^{2} \, F).$$
Using the conservation law (\ref{M1}) we conclude that the density
of energy will be always positive since there exists a minimum
value of the scale factor given by $a_{mim}^{4} = 8 \alpha^{2}
H_{o}^{2}.$ A similar conspiracy occurs in the other extreme where
we approximate $L_{T} \approx  L_{2,3} = L_{2} + L_{3},$ which
shows that the density remains positive definite, since $a(t)$
remains bounded, attaining a maximum in the moment the universe
makes a re-bounce. These extrema occurs precisely at the points
where the total density vanishes. Let us now turn to the generic
conditions needed for the universe to have a bounce and a phase of
accelerated expansion.

\section{Conditions for bouncing and acceleration}

\subsubsection{Acceleration}

From Einstein's equations, the acceleration of the universe is
related to its matter content by \beq 3 \frac{\ddot a}{a} = - \half
(\rho + 3 p). \label{acc} \eeq In order to have an accelerated
universe, matter must satisfy the constraint $(\rho + 3 p)<0$. In
terms of the quantities defined in Eqn. (\ref{NL4}),
\begin{equation}
\rho + 3 p = 2(L-4H^2L_F). \label{TA1}
\end{equation}
 Hence the
constraint $(\rho + 3 p)<0$ translates into \beq L_F >
\frac{L}{4H^2}.\label{23set14} \eeq It follows that any nonlinear
electromagnetic theory that satisfies this inequality yields
accelerated expansion. In our present model it follows that terms
$L_{2}$ and $L_{4}$ produce negative acceleration  and $L_{1}$ and
$L_{3}$ yield inflationary regimes $(\ddot{a} > 0).$

For latter uses we write the value of $\rho + 3p$ for the case of
Lagrangian $L_{T}:$
$$\rho + 3 p = - 6 \alpha^{2} F^{2} + \frac{F}{2} - \frac{6
\mu^{2}}{F} + \frac{10 \beta^{2}}{F^{2}}. $$

\subsubsection{Bouncing}

In order to analyze the conditions for a bouncing it is convenient
to re-write the equation of acceleration using explicitly the
expansion factor $\theta$, which is called the Raychaudhuri
equation:
\begin{equation}
\dot{\theta} + \frac{1}{3} \, \theta^{2} = - \, \half (\rho + 3 p)
\label{4jan2}
\end{equation}
Thus besides condition (\ref{23set14}) for the existence of an
acceleration a bounce needs further restrictions on $a(t)$.
Indeed, the existence of a minimum (or a maximum) for the scale
factor implies that at the bouncing point $B$ the inequality
$(\rho_{B} + 3 p_{B})<0$ (or, respectively, $(\rho_{B} + 3 p_{B})>
0 )$  must be satisfied. Note that at any extremum (maximum or
minimum) of the scale factor the density of energy vanishes. This
is a direct consequence of the first integral of Friedman equation
which, in the Euclidean case, reduces to equation (\ref{29set11}).

\section{Duality on the Magnetic Universe as a consequence of the
inverse symmetry}

The cosmological scenario that is presented here deals with a
cyclic FRW geometry which has a symmetric behavior for small and
big values of the scale factor. This scenario is possible because
the behavior of its energy content at high energy is the same as
it has in its weak regime. This is precisely the case of the
magnetic universe that we are dealing with here. To obtain a
perfect symmetric configuration for our model we will impose a new
dynamical principle:
\begin{itemize}
\item{\textbf{The inverse symmetry principle:}

\vspace{0.5cm}

\emph{ The NLED theory should be invariant under the inverse map
$$ F \rightarrow \tilde{F}= \frac{cte}{F}.$$}}
\end{itemize}

\vspace{0.5cm} For the Lagrangian (\ref{19dez1}), we have chosen the
constant to be $4\mu^2$. This restricts the number of free
parameters from three to two, once a direct application of this
principle implies that $ \beta^{2} = 16 \alpha^{2} \mu^{4}.$  This
symmetry induces a corresponding one for the geometry. Indeed, the
cosmological dynamics is invariant under the associated dual map
\begin{equation}
a(t)  \rightarrow \tilde{a}(t)= \frac{H_{0}}{\sqrt{\mu}} \,
\frac{1}{a} \label{TIS1}
\end{equation}
It is precisely this invariance that is at the origin of the
cyclic property of this cosmological scenario.

Let us point out that the above map is nothing but a conformal transformation.
Indeed, in conformal time, the geometry takes the form

\begin{equation}
ds^{2} = a(\eta)^{2} \left( d\eta^{2} -  dr^{2} - r^{2} d\Omega^{2}
\right). \label{6out730}
\end{equation}
Thus making the conformal map
$$ \tilde{g}_{\mu\nu} = \omega^{2} \, g_{\mu\nu}$$
where $\omega = \lambda /a^{2},$ and $\lambda \equiv
H_{0}/\sqrt{\mu}.$ Note that although the Lagrangian $L_{T}$ is not
invariant under a conformal transformation, the average procedure
used to make compatible the dynamical system of the electomagnetic
field and the Friedman equation is invariant. Indeed, we have
$$ \tilde{F} = \tilde{g}^{\mu\nu} \tilde{g}^{\alpha\beta} \,
F_{\alpha\mu} F_{\beta\nu} =  \frac{4 \mu^{2}}{F}.$$

\section{A complete scenario}

Electromagnetic radiation described by a
maxwellian distribution has driven the geometry of the universe for a
period. Let us now analyze the modifications introduced
by the non linear terms in the cosmic scenario. The simplest way
to do this is to combine the previous lagrangian with the
dependence of the magnetic field on the scale factor. We set
\begin{equation}
L_{T} = \alpha^{2} \,F^2 -\frac{1}{4}\,F   - \frac{\mu^{2}}{F} +
\frac{\beta^{2}}{F^{2}} \label{11jan1}
\end{equation}
where $\beta$ is related to the other parameters $\alpha$ and
$\mu$ by the inverse symmetry principle, as displayed above.

\subsection{Potential}
It will be more direct to examine the effects of the magnetic
universe controlled by the above lagrangian if we undertake a
qualitative analysis using an analogy with classical mechanics.
Friedman's equation reduces to the set

\beq \dot a^2+V(a) = 0 \label{11feve} \eeq where \beq
V(a)=\frac{A}{a^6}-\frac{B}{a^2}-Ca^6 + D a^{10} \label{pot} \eeq is
a potential that restricts the motion of the localization $a(t)$ of
the ``particle''. The constants in $V$ are given by
$$
A=\frac{4 \alpha^{2} H_0^4}{3},\;\;\;\;
B=\frac{H_0^2}{6},\;\;\;\;C=\frac{\mu^2}{6
H_0^2},\;\;\;\;D=\frac{4 \alpha^{2} \mu^4}{3 H_0^4},
$$
and are positive.

 The dependence of the field as $H = H_{0}/a^{2}$ implies the existence
of four distinct epochs, which we will analyze now.

 The derivative $dL/dF$ has three zeros, in which $\rho + p$ vanishes.  In the
case of pure magnetic universe the value of $F$ is always positive.
We distinguish four eras.

\subsection{The four eras of the Magnetic Universe} \label{dynamics}

 The dynamics of the universe with matter density
given by Eqn.($29$) can be obtained qualitatively from the analysis
of Einstein's equations. We distinguish four distinct periods
according to the dominance of each term of the energy density. The
early regime (driven by the $F^{2}$ term); the radiation era (where
the equation of state $p = 1/3\rho$ controls the expansion); the
third accelerated evolution (where the 1/F term is the most
important one) and finally the last era where the $1/F^{2}$
dominates and in which the expansion stops, the universe re-bounces
and enters in a collapse era.

\subsubsection{Bouncing era}

In the strong field limit the value of the scalar of curvature is
small and the volume of the universe attains its minimum, the
density of energy and the pressure are dominated by the terms
coming from the quadratic lagrangian $F^{2}$ and is approximated
by the forms
\begin{eqnarray}
\rho & \approx & \frac{H^{2}}{2} \, ( 1 - 8 \alpha^{2}\, H^{2})
\nonumber \\
p & \approx & \frac{H^{2}}{6} \, ( 1 - 40 \alpha^{2}\, H^{2})
\label{2Maio11}
\end{eqnarray}

Using the dependence $ H= H_o/a^2,$ on equation (\ref{11feve}) leads
to
\begin{equation}
\label{eqA2} \dot{a}^2=\frac{kH_o^2}{6\,a^2}
\left(1-\frac{8\alpha^{2} H_o^2}{a^4}\right)-\epsilon.
\end{equation}
We remind the reader that we limit our analysis here to the
Euclidean section $(\epsilon = 0).$ As long as the right-hand side
of equation (\ref{eqA2}) must not be negative it follows that, the
scale-factor $a(t)$ cannot be arbitrarily small. Indeed, a
solution of (\ref{eqA2}) is given as
\begin{equation}
\label{A(t)} a^2 = H_{o} \,\sqrt{\frac{2}{3} \,( t^2
+12\,\alpha^{2})}.
\end{equation}
 The radiation period can be
achieved from the above equation by setting $\alpha=0$. As a
consequence the average strength of the magnetic field $H$ evolves
in time as
\begin{equation}
 H^2 = \frac{3}{2}\,\frac{1}{{ t^2} +
12\,\alpha^{2}}.\protect\label{H(t)}
\end{equation}
 Note that at $t = 0$ the
radius of the universe attains a minimum value at the bounce:
\begin{equation}
\label{Amin} a^2_{B} = H_{o} \, \sqrt{8\,\alpha^{2}}.
\end{equation}
Therefore, the actual value of $a_{B}$ depends on $H_o$, which -
for given $\alpha, \, \mu$ turns out to be the sole free parameter
of the model. The energy density $\rho$ reaches its maximum for
the value $\rho_{B}=1/64\alpha^{2}$
 at the instant $t=t_B$, where
\begin{equation}
 t_{B} = \sqrt{12\,\alpha^{2}} .\label{tc}
\end{equation}
For smaller values of $t$ the energy density decreases, vanishing
at $t = 0$, while the pressure becomes negative.
Only for very small times $ t \, <  \, \sqrt{4 \alpha^{2} / k}$
 the non-linear
effects are relevant for cosmological solution of the normalized
scale-factor. Indeed, solution (\ref{A(t)}) fits the standard
expression of the Maxwell case at the limit of large times.

\subsubsection{Radiation era} The standard Maxwellian term dominates
in the intermediary regime. Due to the dependence on $a^{-2}$ of the
field, this phase is defined by $H^{2} >> H^{4}$ yielding the
approximation
\begin{eqnarray}
\rho & \approx & \frac{H^{2}}{2} \nonumber \\
p & \approx & \frac{H^{2}}{6} \, \label{3Maio11}
\end{eqnarray}
This is the phase dominated by the linear regime of the
electromagnetic field. Its properties are the same as described in
the standard cosmological model.

\subsubsection{The accelerated era: weak field drives the
cosmological geometry}

 When the universe becomes larger, negative powers of $F$
 dominates and the distribution of energy becomes typical of an
accelerated universe, that is:
\begin{eqnarray}
\rho & \approx & \frac{1}{2} \, \frac{\mu^{2}}{H^{2}}
\nonumber \\
p & \approx & \frac{- 7}{6} \, \frac{\mu^{2}}{H^{2}} \label{4Maio11}
\end{eqnarray}

  In the intermediate regime between the radiation and the acceleration regime
  the energy content is described by the combined form
$$
\rho = \frac{H^2}{2} + \frac{\mu^2}{2} \frac{1}{ H^2},
$$
or, in terms of the scale factor,
  \beq \rho = \frac{H_0^2}{2}\;
\frac{1}{a^4} + \frac{\mu^2}{2H_0^2}\; a^4. \label{density} \eeq

 For small $a$ it is the ordinary
radiation term that dominates. The $1/F$ term takes over only
after $a=\sqrt{H_0}/\mu$, and would grows without bound
afterwards. In fact, the curvature scalar is
$$
R = T^\mu_{\;\mu} = \rho-3p = \frac{4\mu^2}{H_0^2}\; a^4,
$$
showing that one could expect a curvature singularity in the
future of the universe for which $a\rightarrow\infty.$  We shall
see, however that the presence of the term $1/F^{2}$ changes this
behavior.

Using this matter density in Eqn.(\ref{acc}) gives
$$
3 \frac{\ddot a}{a} +  \frac{H_0^2}{2}\; \frac{1}{a^4} - \frac 3 2
\frac{{\mu^2}}{H_0^2}\; a^4=0.
$$
To get a regime of accelerated expansion, we must have
$$
 \frac{H_0^2}{a^4} -  3
\frac{{\mu^2}}{H_0^2}\; a^4 < 0,
$$
which implies that the universe will accelerate for $a>a_c$, with
$$
a_c = \left(\frac{H_0^4}{3\mu^2}\right)^{1/8}.
$$

\subsubsection{Re-Bouncing}

For very big values of the scale factor the density of energy can
be approximated by
\begin{equation}
\rho \approx \frac{\mu^{2}}{F} - \frac{\beta^{2}}{F^{2}}
\end{equation}
and we pass from an accelerated regime to a phase in which the
acceleration is negative. When the field attains the value $F_{RB}
= 16 \alpha^{2} \mu^{2}$ the universe changes its expansion to a
collapse. The scale factor attains its maximum value
$$a^{4}_{max}
\approx \frac{H_{0}^{2}}{8 \alpha^{2} \mu^{2}}.$$

\section{Positivity of the density of energy}
The total density of energy of the BUC is always positive definite
(see equation \ref{29set11}). In the bouncing and in the re-bouncing
eras it takes the value $\rho_{B} = \rho_{RB} = 0.$ At these points
the density is an extremum. Actually, both points are minimum of the
density. This is a direct consequence of equations (\ref{M1}) and
(\ref{29set11}). Indeed, derivative of (\ref{M1}) at the bouncing
and at the re-bouncing yields
$$ \ddot{\rho}_{B} = \frac{3}{2} \, p_{B}^{2} > 0.  $$
Thus there must exists another extremum of $\rho$ which should be
a maximum. This is indeed the case since there exists a value on
the domain of the evolution of the universe between the two minima
such that
$$ \rho_{c} + p_{c} = 0.$$  At this point we have

$$ \ddot{\rho} + \dot{p}_{c} \, \theta_{c} = 0$$
showing that at this point $c$ the density takes its maximum
value.

\section{The behavior of the scale factor}
Let us pause for a while and describe the form of the scale factor
as function of time in the four regimes. To simplify such
description let us separate in three parts:
\begin{description}
\item{Phase A: Bouncing-Radiation}
\item{Phase B: Radiation-Acceleration}
\item{Phase C: Acceleration-Re-bouncing}
\end{description}
characterized respectively by the dynamics controlled by:  $L_{A} =
 L_{1} + L_{2}; L_{B} =  L_{2} + L_{3}; L_{C} = L_{3} + L_{4}.$
 It is straightforward to obtain an analytical
expression for each one of these periods, which can be analytically
continued through distinct eras. We obtain for phase $A:$
\begin{equation} a_{BR}(t)=
\left[\frac{2}{3}H_{0}^2\left[\left(t-t_1\right)^2+12\alpha^2\right]\right]^{1/4}\label{27out17}
\end{equation}
The inverse symmetric phase $C$ is given by
\begin{equation}
a_{ARB}(t)=\left[\frac{2}{3}\frac{\mu^2}{H_{0}^2}\left[\left(t-t_3\right)^2+12\alpha^2\right]\right]^{-1/4}
\end{equation}
and for the intermediary phase $B$, we have:
\begin{equation}
a_{RA}(t)= {\left[\frac{H_{0}^2}{\mu}
\frac{1-cos\left(Jsn\left(2\sqrt{\frac{2\mu}{3}}\left(t-t_2\right),\frac{\sqrt{2}}{2}\right)\right)}
{1-cos\left(Jsn\left(2\sqrt{\frac{2\mu}{3}}\left(t-t_2\right),\frac{\sqrt{2}}{2}\right)\right)}\right]}^{1/4}
\label{ana}
\end{equation}
where $Jsn$ is the inverse of a first kind elliptic function, or the
Jacobi function $JacobiSN$. In order to express the analytical
continuation of the scalar factor trough the different eras it is
convenient to re-write equation (\ref{ana}) in the inverse form:
\begin{equation}
t - t_{2} = \sqrt{\frac{3}{\mu}} EllipticF \left[ \arccos
\left(\frac{H_{0}^{2} - \mu a^{4}}{ H_{0}^{2} + \mu a^{4}}\right) ,
\frac{\sqrt{2}}{2} \right] \label{ana2}.
\end{equation}
In the limit $ \mu a^{4} \ll H_{0}^{2}$ expression (\ref{ana2})
becomes
$$ t - t_{2} = \sqrt{\frac{3}{2 H_{0}^{2}}} \, \, a^{2}.$$

Now, in the limit $ t - t_{1} \gg 12 \alpha^{2} $ expression $
a_{ARB}(t)$ reduces to
$$ a(t) =\left[\frac{2}{3}H_{0}^2\left[\left(t-t_1\right)^2
\right]\right]^{1/4} $$ corresponding to the radiation era, which is
continued in form $ a_{RA}(t)$ above by identifying $ t_{1} =
t_{2}.$

On the opposite case  $ \mu a^{4} \gg H_{0}^{2}$ expression
(\ref{ana2}) goes to

$$ t - t_{2} = \sqrt{\frac{3}{2 \mu}} \left( \pi - \sqrt{H_{0}^{2}}{\mu} \, \frac{1}{a^{2}} \right),$$
that is

$$ a(t) = \left[\frac{2}{3}\frac{\mu^2}{H_{0}^2}\left[\left(t-t_2\right)^2\right]\right]^{-1/4}
$$
which is continued analytically to the phase $C$ by the
identification $ t_{3} = t_{1} + \pi \sqrt{\frac{3}{2 \mu}}.$

\section{Conclusion}
The simplified toy model presented here displays many
regular properties that should be worth of further investigation. In
particular, it provides a spatially homogeneous and isotropic FRW
geometry which has no Big Bang and no Big Rip. It describes
correctly the radiation era and allows for an accelerated phase
without introducing any extra source.

The particular form of the dynamics of the magnetic field is
dictated by the inverse principle, which states that the behavior of
the field is invariant under the mapping $ F \rightarrow \tilde{F}=
\frac{4 \mu^{2}}{F}.$ This reflects on the symmetric behavior of the
geometry by the dual map $ a \rightarrow \tilde{a}= H_{0}/\sqrt{\mu}
\, a.$

The particular form of NLED is based on the principle that shows an intimate relation between strong and weak
field configurations. This inverse-symmetry principle reduces the
number of arbitrary parameters of the theory and allows for the
regular properties of the cosmical model. The universe is a cyclic
one, having its main characteristics synthesized in the following
steps:
\begin{itemize}
\item{Step 1: The universe contains a collapsing phase in which the scalar factor
attains a  minimum value $a_{B}(t);$}
\item{Step 2: after the bouncing the universe expands with $\ddot{a} <0;$}
\item{Step 3: when the $1/F$ factor dominates the universe enters
an accelerated regime;}
\item{Step 4: when $1/F^{2}$ dominates the acceleration changes the sign and starts a phase
in which $\ddot{a} <0$ once more, the scale factor attains a
maximum and re-bounces starting a new collapsing phase;}
\item{Step 5: the universe repeats the same behavior passing
steps 1, 2, 3 and 4 again and again, indefinitely.}
\end{itemize}
\subsection*{Appendix A: Static and spherically symmetric
electromagnetic solution and the asymptotic regime}

We have made an analysis of the modification of electrodynamics in a
cosmological context. We are not arguing that these effects are more
than the response of the universe to local electrodynamics
properties. Some decades ago Wheeler and Feynman made a conjecture
that local properties of electrodynamics (e.g. the Lienard-Wiechert
potential) may just be a consequence of such cosmic response
inducing the elimination of advanced fields. However, if one takes
these modifications as local change of electrodynamics, we should
check consistency of the theory with conventional electromagnetism.
We shall restrict ourselves here to the case of the static electric
field generated by a point charged particle. For a general nonlinear
Lagrangian $L=L(F)$, the EOM for the point charge reduces to
$$
r^2 L_F\;E(r)= {\rm const.}
$$
In the case of the Lagrangian given in Eqn.(\ref{lag}) we get
\beq
-\frac{1}{4 E^{5}(r)} \left( 16 \alpha^2 E(r)^8 + E(r)^{6} - \mu^{2}
E(r)^{2} - \beta^{2}  \right) = \frac{q}{r^2}. \label{ef} \eeq

The polynomial in $E$ that follows from this equation cannot be
solved exactly, but to study the dependence of $E$ with $r$ we can
plot from (\ref{ef}) the function $r=r(E)$.

    It also follows from the
plot that $E\rightarrow$ constant for $r\rightarrow\infty$.

By taking derivatives of Eqn.(\ref{ef}) it can be shown that the
function $E(r)$ has no extrema \footnote{Note that $E=0$ is not a
solution of Eqn.(\ref{ef}).}. Hence, the modulus of the electric
field decreases monotonically with increasing $r$, from an
infinite value at the origin to a constant (nonzero but small)
value at infinity. Eqn.(\ref{ef}) then shows that $E_\infty =
constant \neq 0$. This situation is akin to that in the theory
defined by the action
$$
S =\frac{M_{\rm P}^2}{2} \int \sqrt{-g} \left( R -
\frac{\alpha^4}{ R} \right) d^4x.
$$
It was shown in \cite{carroll1} that the static and spherically
symmetric solution of this theory does not approach Minkowski
asymptotically; it tends instead to (anti)-de Sitter space-time. We
shall see that a similar situation occurs in the case of NLED.

Regarding the behavior of the field for small values of $r$, if we
compare the term corresponding to Maxwell's case in Eqn.(\ref{ef})
with the other two, we get that for the field to be Maxwell-like
there are conditions on the value of the free parameters to be
fulfilled, thats is:
$$ E^2 << \frac{1}{16\alpha^2}.
$$

$$ \mu^{2}  << E^4 $$
$$ \beta^{2} << E^6.$$

With the explicit dependence for the field given by $E(r)=q/r^2$, it
would be possible to set a value for $\alpha$ in agreement with the
observation by
$$
\alpha^2<< \frac{r_0^4}{16q^2},
$$ and analogous expressions for $\mu$ and $\beta,$
where $r_0$ is a reference value set by the experiment. In the
case we fix the value of $\beta$ by the inverse symmetry, the
condition on $\beta$ reduces to the other two.

\subsubsection{Asymptotic regime}

Let us make an extra comment on the above case of a point charge
particle at spatial infinity. The energy-momentum tensor has the
form:
\begin{equation}
T_{\mu\nu} = - L \eta_{\mu\nu} - 4 L_{F} \, F_{\mu\alpha} \,
F^{\alpha}_{}{\nu} \label{21junho7}
\end{equation}
which in the present case $F_{01} = E(r)$ is
\begin{eqnarray}
T^{0}_{0}= T^{1}_{1} &=& \frac{1}{4E^{4}} \, \left( 48 \alpha^{2}
E^{8} + 2 E^{6} - 6 \mu^{2} E^{2} - 5 \beta^{2}
\right)\nonumber \\
 T^{2}_{2}= T^{3}_{3} &=& - \, \frac{1}{4E^{4}} \, \left( 16 \alpha^{2}
E^{8} + 2 E^{6} +2  \mu^{2} E^{2}  -  \beta^{2} \right)
\label{21junho10}
\end{eqnarray}
In the asymptotic regime, we can set $ E = E_{\infty}= X $ and,
for the energy-momentum tensor one obtains
\begin{equation}
T^{0}_{0}= T^{1}_{1} =  T^{2}_{2}= T^{3}_{3} = - \,\frac{1}{4 X^{2}}
\, \left( X^{3} + 3 \mu^{2} X + 2 \beta^{2} \right)
\label{21junho11}
\end{equation}
which mimics a $\Lambda$ term.

If we add an extra term in the Lagrangian we could eliminate the
residual constant field at infinity. In the case of Maxwell
Electrodynamics such ambiguity of choice does not arise due to its
linearity. However, for non-linear electromagnetic theory a new
possibility occurs which concerns the geometrical structure at
infinity. This means that for the non-linear electrodynamics the
fact that at infinity the field is a constant does not implies that
it vanishes. Such property can be translated in a formal question,
that is,  what is the asymptotic regime of the geometry of
space-time: Minkowski or de Sitter? .

In classical linear Electrodynamics the answer to that question
was known and did not pose any ambiguity. No longer so if
non-linear electromagnetic field is combined with the equations of
general relativity. The possibility of the de Sitter structure
must be considered. In theories in which a solution distinct from
zero for the equation $L_{F} = 0$ exists, such a question has to
be investigated combined with cosmology. In a recent paper
\cite{Novello-Neves} a new look into this question was considered
by the exam of a proposed relation of the apparent mass of the
graviton and the cosmological constant. We will come back to this
question elsewhere.

\subsection*{Appendix B: The fundamental state} A simple look into
the equation of motion in NLED shows the existence of a very
particular solution such that its energy distribution is the same
as the one in the vacuum fundamental state represented by an
effective cosmological constant. Indeed, the eom is given by
\begin{equation}
(L_{F} F^{\mu\nu})_{;\nu} = 0.
\end{equation}
Consider the particular solution $F = F_{0} = constant$ such that
$$ 2 \alpha^{2} F^{4} - \frac{F^{3}}{4} + \mu^{2} F - 2 \beta^{2}
= 0.$$ This is the condition that satisfies the equation of motion
since $L_{F}$  vanishes at this value  $F_{0}.$ In this state the
corresponding energy-momentum takes the form
$$ T_{\mu\nu} = \Lambda \, g_{\mu\nu}, $$
where
$$ \Lambda = - L (F_{0}) = \frac{1}{F_{0}^{2}} \left(
\frac{1}{8} \, F_{0}^{3} + \frac{3\mu^{2}}{2} \, F_{0} - 2
\beta^{2} \right). $$ This property is typical of NLED since there
is no possibility of the linear Maxwell theory having such
particular solution.

\subsection*{Appendix C: Causality} Most\footnote{The authors thanks
SEP Bergliaffa for using this section which was made in
collaboration with him.}  of our description of the universe is
based on the behavior of light in a gravitational field. In a FLRW
scenario the existence of a horizon inhibiting the complete
interchange of information between arbitrary parts of the universe
associated with the observation of the high degree of isotropy of
the CBR generated a causal difficulty: different parts of the
universe - in the standard FLRW geometry could not have enough
time to homogeneize. The inflationary scenario had its appeal
precisely for the solution it brought to this problem.

Non linear theories of Electrodynamics presented a completely new
look into the causal properties, which we will now overview very
briefly. \subsubsection{Causal Structure of Nonlinear
Electrodynamics}

 The main lesson we can extract from the analysis of the
propagation of the wave front in nonlinear theory is contained in
a theorem that can be demonstrated using Hadamard method to deal
with the discontinuity of the field and which can be synthesized
in a single sentence dealing with the modification of the
geometry,  generating an effective metric that controls the
properties of the space. In order to show this let us make a very
short summary of it.

\subsubsection{The Effective Metric} \label{intro}

We decided to give a very short review of the propagation of the
electromagnetic waves in the NLED, since it has very peculiar
properties that modify the standard description in the linear
Maxwellian case. The reader interested in more details can consult
 \cite{Salim}.

Historically, the first example of the idea of effective metric
was presented in 1923 by W. Gordon. In modern language, the wave
equation for the propagation of light in a moving non-dispersive
medium, with slowly varying refractive index $n$ and 4-velocity
$u^\mu$:
$$
\left[ \partial_\alpha \partial^\alpha + (n^2-1) (u^\alpha
\partial_\alpha)^2 \right] F_{\mu\nu} = 0.
$$
Taking the geometrical optics limit, the Hamilton-Jacobi equation
for light rays can be written as  $ g^{\mu\nu}k_\mu k_\nu = 0 $ (see
\cite{Salim} for details), where \beq g^{\mu\nu}= \eta^{\mu\nu}+
(n^2 - 1) u^\mu u^\nu \label{gordonm} \eeq is the effective metric
for this problem. It must be noted that only photons in the
geometric optics approximation move on geodesics of $g^{\mu\nu}$:
the particles that compose the fluid couple instead to the
background Minkowskian metric. Let us now study in detail the
example of nonlinear electromagnetism. We start with the action \beq
S = \int \sqrt{-\gamma}\; L(F) \;d^{4}x ,\protect \label{N1}
\end{equation}
where $F \equiv F^{\mu\nu}F_{\mu\nu}$  and $L$ is an arbitrary
function of $F$. Notice that $\gamma$ is the determinant of the
background metric. which we take in the following to be that of flat
spacetime, but the same techniques can be applied when the
background is curved. Varying this action w.r.t. the potential
$W_{\mu}$, related to the field by the expression
$$
F_{\mu\nu} = W_{\mu;\nu} - W_{\nu;\mu} = W_{\mu,\nu} -
W_{\nu,\mu},
$$
we obtain the Euler-Lagrange equations of motion (EOM)
\begin{equation}
(L_{F} F^{\mu\nu})_{;\nu} = 0,
 \protect\label{N2}
\end{equation}
where $L_{F}$ is the derivative $ L_{F} \equiv   \partial L /
\partial F.$ In the particular case of a linear dependence of the
Lagrangian with the invariant $F$ we recover Maxwell's equations
of motion.

As mentioned in the Introduction, we want to study the behaviour
of perturbations of these EOM around a fixed background solution.
Instead of using the traditional perturbation method, we shall use
a more elegant method set out by Hadamard. In this method, the
propagation of low-energy photons are studied by following the
evolution of the wave front, through which the field is continuous
but its first derivative is not. To be specific, let $\Sigma$ be
the surface of discontinuity defined by the equation
$$\Sigma(x^{\mu}) = constant.$$
The discontinuity of a function $J$ through the surface $\Sigma$
will be represented by $[J]_\Sigma$, and its definition is
$$
[J]_\Sigma \equiv \lim_{\delta\rightarrow 0^+} \left( \left.
J\right|_{\Sigma +\delta} - \left. J \right|_{\Sigma -
\delta}\right) .
$$
The discontinuities of the field and its first derivative are
given by
\begin{equation}
[F_{\mu\nu}]_{\Sigma} = 0,
\;\;\;\;\;\;\;\;\;\;\;\;\;\;\;\;\;\;\;\;\;\;
[F_{\mu\nu,\lambda}]_{\Sigma} = f_{\mu\nu}  k_{\lambda},
\protect\label{N8}
\end{equation}
where the vector $k_{\lambda}$ is nothing but the normal to the
surface $\Sigma$, that is, $k_{\lambda} = \Sigma_{,\lambda},$ and
 $f_{\mu\nu}$ represents the discontinuity of the field.

  To set the stage for the nonlinear case, let us
first discuss the propagation in Maxwell's electrodynamics, for
which $ L_{FF} = 0$. The EOM then reduces to $F^{\mu\nu}_{;\nu} =
0$, and taking the discontinuity we get
\begin{equation}
f^{\mu\nu}  k_{\nu} = 0. \protect\label{N10}
\end{equation}
The other Maxwell equation is given by ${F^{*}_{\mu\nu}}^{;\nu} =
0$ or equivalently,
\begin{equation}
F_{\mu\nu;\lambda} + F_{\nu\lambda;\mu} + F_{\lambda\mu;\nu} = 0.
\protect\label{N12}
\end{equation}
The discontinuity of this equation yields
\begin{equation}
f_{\mu\nu}  k_{\lambda} + f_{\nu\lambda}  k_{\mu} + f_{\lambda\mu}
k_{\nu} = 0. \protect\label{N13}
\end{equation}
Multiplying this equation by $k^{\lambda}$ gives
\begin{equation}
f_{\mu\nu}  k^2 + f_{\nu\lambda}  k^{\lambda} k_{\mu} +
f_{\lambda\mu} k^{\lambda} k_{\nu} = 0, \protect\label{N14}
\end{equation}
where $k^2 \equiv k_{\mu} k_{\nu} \gamma^{\mu\nu}. $ Using the
orthogonality condition from Eqn.(\ref{N10}) it follows that
\begin{equation}
f^{\mu\nu}  k^2  = 0 \protect\label{N15}
\end{equation}
Since the tensor associated to the discontinuity cannot vanish (we
are assuming that there is a true discontinuity!) we conclude that
the surface of discontinuity is null w.r.t. the metric
$\gamma^{\mu\nu}$. That is,
\begin{equation}
k_{\mu} k_{\nu} \gamma^{\mu\nu} = 0. \protect\label{N16}
\end{equation}
It follows that $ k_{\lambda;\mu} k^{\lambda} = 0$, and since the
vector of discontinuity is a gradient,
\begin{equation}
k_{\mu;\lambda} k^{\lambda} = 0. \protect\label{N17}
\end{equation}
This shows that the propagation of discontinuities of the
electromagnetic field, in the case of Maxwell's equations (which
are linear), is
along the null geodesics of the Minkowski background metric.\\[.1cm]

Let us apply the same technique to the case of a nonlinear
Lagrangian for the electromagnetic field, given by $L(F)$. Taking
the discontinuity of the EOM, Eqn.(\ref{N2}), we get
\begin{equation}
L_{F} f^{\mu\nu} k_{\nu} + 2 \eta\;L_{FF}\;  F^{\mu\nu} k_{\nu} =
0, \protect\label{N18}
\end{equation}
where we defined the quantity $\eta$ by $
F^{\alpha\beta}f_{\alpha\beta} \equiv \eta. $ Note that contrary
to the linear case in which the discontinuity tensor $f_{\mu\nu}$
is orthogonal to the propagation vector $k^{\mu}$, here there is a
complicated relation between the vector  $f^{\mu\nu} k_{\nu}$ and
quantities dependent on the background field. This is the origin
of a more involved expression for the evolution of the
discontinuity vector, as we shall see next. Multiplying equation
(\ref{N14}) by $F^{\mu\nu}$ we obtain
\begin{equation}
\eta  \;k^2 + F^{\mu\nu} f_{\nu\lambda}  k^{\lambda} k_{\mu} +
F^{\mu\nu} f_{\lambda\mu}  k^{\lambda} k_{\nu} = 0.
\protect\label{N141}
\end{equation}
Now we substitute in this equation the term $  f^{\mu\nu} k_{\nu}
$ from Eqn.(\ref{N18}), and we arrive at the expression
\begin{equation}
L_{F} \, \eta k^2 - 2 \,  L_{FF} \eta ( F^{\mu\lambda} k_{\mu}
k_{\lambda} -
 F^{\lambda\mu} k_{\mu} k_{\lambda} ),
\protect\label{N19}
\end{equation}
which can be written as $g^{\mu\nu}k_\mu k_\nu = 0$, where
\begin{equation}
g^{\mu\nu}= L_{F} \, \gamma^{\mu\nu} - 4 \, L_{FF}\, F^{\mu\alpha}
F_{\alpha}{}^{\nu}. \protect\label{N20}
\end{equation}
We then conclude that the high-energy photons of a {\em nonlinear}
theory of electrodynamics with $L=L(F)$ do not propagate on the
null cones of the background metric but on the null cones of an
{\em effective} metric, generated by the self-interaction of the
electromagnetic field.

\subsubsection{Causal properties in the fundamental state} Let us
look for the causal structure in the case the electromagnetic
field rests on its fundamental state. From the calculation made in
the previous  chapter the photons propagate in an effective
geometry which is given by equation (\ref{N20}). In the case of
the inverse symmetry Lagrangian the effective metric tensor takes
the form:
\begin{equation}
g^{\mu\nu}_{(eff)}= (- \, \frac{1}{4} + \frac{\mu^{8}}{F^{2}}) \,
g^{\mu\nu}
 +  \frac{8 \mu^{8}}{F^{3}} \,  F^{\mu\alpha} F_{\alpha}{}^{\nu}.
\label{CP2}
\end{equation}
The fundamental state is the particular solution in which
$$ F^{2} = 4 \mu^{8},  $$
which corresponds to an energy-momentum tensor equivalent to a
fluid distribution characterized by the condition $ \rho + p = 0$
and generates a deSitter geometry for the background metric
$g_{\mu\nu}$ as seen by all forms of matter and energy content -
as far as we neglect the gravitational influence of such remaining
matter and energy. However, from the above calculation, we
conclude that the photons do not propagate in such deSitter space
but instead in an effective metric which is provided by the form:
\begin{equation}
g^{\mu\nu}_{(eff)} = \pm \frac{1}{\mu^{4}} \,  F^{\mu\alpha}
F_{\alpha}{}^{\nu}. \label{CP3}
\end{equation}
This is a very peculiar and interesting situation that can be
described by the following sentence:

\begin{itemize}

\item{ \emph{\textbf{ The
fundamental state of the theory described by the inverse symmetric
Lagrangian generates a deSitter universe felt by all existing
matter with one exception: the photons which follow geodesics in
the above anisotropic geometry $g^{\mu\nu}_{(eff)}.$}}}

\end{itemize}

\section*{Acknowledgements}

JS is supported by CNPq. MN acknowledges support of FAPERJ and
CNPq. ANA is supported by CAPES.

\end{document}